\documentclass{appolb}
\usepackage{epsfig}

\begin{document}
\newcount\eLiNe\eLiNe=\inputlineno\advance\eLiNe by -1
\title{THE NUCLEAR PIONS AND QUARK DISTRIBUTIONS\\
IN DEEP INELASTIC SCATTERING ON NUCLEI\thanks{Presented by J.
Ro\.zynek (rozynek@fuw.edu.pl) at the XXVII Mazurian Lakes School of
Physics {\it Growth Points of Nuclear Physics A.D. 2001}, September
2-9, 2001, Krzy\.ze, Poland.}
}
\author{J. Ro\.zynek, G.Wilk
\address{ The Andrzej So\l tan Institute for Nuclear Studies, Warsaw,
Poland}}
\maketitle

\begin{abstract}
We propose simple Monte Carlo method for calculating parton
distribution in nuclei. Only events satisfying the exact kinematical
constrains of the corresponding deep-inelastic reaction probing given
nuclear distribution are selected to form the final distribution we
are looking for. The EMC effect is automatically included by means of
two parameters, which characterize the change of the nuclear pion
field. Good agreement with experimental data in the broad range of
variable $x$ is obtained.
\end{abstract}

\section{Introduction}
The study of partonic distributions inside the nucleon and nuclei has
already long history \cite{History}. Here we shall describe simple
Monte Carlo method for calculating parton distributions in nuclei
with special emphasis on the observed differences between such
distributions for free and nuclear nucleons (known under the name
EMC-effect) \cite{RW}. The parton picture of nucleon was originally 
formulated in the infinite momentum frame \cite{Bj}. However, nuclear
effects are more visible in the nucleon (or nuclear) target rest
frame. Such change of frames has profound dynamical consequences.
Whereas in the infinite momentum frame partons can be treated as
on-shell objects (with some small current masses), in the rest frame
they are dressed, far off-shell objects with masses consisting
substantial part of the nucleon mass. The important point is how
to find in this frame the proper parton energy. The nucleon is no 
longer contracted by Lorentz transformation and the corresponding
interaction picture is complicated one. The nuclear parton density
distributions are therefore obtained by generating initial parton
momenta in nucleons and calculating the corresponding light-cone
longitudinal momentum fractions, the so called Bjorken $x=x_B$,
which must be identical in both frames.

Our approach is based on the model where valence parton momenta in
hadron at rest are calculated from a spherically symmetric Gaussian
distribution with a width derived from the Heisenberg uncertainty
relation, whereas the sea parton contributions result from similar
gaussian distribution but with a width dictated by the presence of
virtual pions in hadron \cite{RW}. When going to the nuclear case
these initial gaussian momentum distributions are changed accordingly
in order to account for the presence of nuclear medium (like
rescattering effects or changes in the virtual pion clouds in the
nuclear matter). The energy momentum conservation is always strictly
imposed and plays vital role in getting our results. The nuclear
parton density distributions are then obtained by generating initial
parton momenta of nucleons and calculating the corresponding
light-cone longitudinal momentum fractions $x_B$.

\section{Deep Inelastic Scattering}

We start with short recollection of necessary theoretical points. In
the deep inelastic electron-nucleon scattering the cross section is
given by:
\begin{equation}
  d\sigma=L_{\mu\nu}W^{\mu\nu} \label{eq:xsection}
\end{equation}
where the hadronic tensor $W^{\mu\nu}$ can be expressed in terms
of electromagnetic currents as:
\begin{equation}
  W_{\mu\nu}=\int d^{4}p \exp(iq\xi) <p \mid J_{\mu} J_{\nu}(0) \mid p>.
  \label{eq:tensor}
\end{equation}
The spatial variable $\xi$ is directly connected to the correlation
length existing in this process. The virtual foton momentum transfer
is given by:
\begin{equation}
  q=(\nu,0,0,-\sqrt{\nu^{2}+Q^{2}}) . \label{eq:gamma}
\end{equation}
In the Bjorken limit $Q^{2}\rightarrow\infty$ the $x=Q^{2}/2M\nu\approx
j^{+}/p^{+}$ is fixed and $q^{2}/\nu^{2}\rightarrow 0$. In this limit
$q^{-}=q^{0}-q^{3}\rightarrow\infty$ but $q^{+}=-Mx/\sqrt{2}$
remains finite. These imply $\xi^{+}\rightarrow0$ and
$\xi^{-}\unlhd\sqrt{2}/Mx$. Alltogether this gives the following
restrictions for $\xi$:
\begin{equation}
  \xi_{0}\leq1/Mx,\qquad \xi_{z}\equiv z\leq1/Mx . \label{eq:restr}
\end{equation}
We have therefore two resolutions scales in deep inelastic
scattering:
\begin{itemize}
\item[$(i)$] $1/\sqrt{Q^{2}}$ connected with virtuality of
$\gamma$ probe. Any two different $Q^{2}$ resolutions are connected
via well known A-P evolution equation \cite{AP}.
\item[$(ii)$] $\frac{1}{Mx}  = z$ being distance how far can propagate
the anti-quark in the medium, see Fig.1. Notice that small $x$ means a
relatively large correlation length $z$. Because final state quark
interaction within the nuclear envirovment is practically not known
the small $x$ region opens room for different phenomenological
models and in this paper we shall propose a new mechanism for the
nuclear shadowing.
\end{itemize}

Deep inelastic scattering of electrons on nuclear targets can be
regarded as two step process: at first nucleus is replaced by
composition of nucleons and (effective) pions representing quanta of
nuclear binding forces, then impinging electrons interact with partons
(quarks) composing those nucleons and pions. Formally it means that
nuclear partonic distribution (structure function) $F_2^A$ can be
written as convolution,
\begin{equation}
 F^{A}_{2}(x_{A})/x_{A} = A \int \int dy_{A} dx \delta
(x_{A}-y_{A}x) \rho ^{A}(y_{A}) F^{N}_{2}(x)/x, \label{eq:F2}
\end{equation}
of nucleon distribution function in the nucleus, $\rho^{A}(y_A)$, and
structure function of free nucleon, $F^{N}_{2}(x)$. The $x_{A}/A$ is
the ratio of quark and the nucleus longitudinal momenta (i.e., it is
the Bjorken variable for the nucleus) whereas
$y_{A}/A=p^{+}/P^{+}_{A}$ is the ratio of the nucleon and nuclear
longitudinal momenta. Finally $ x$ is the ratio of the quark 
and nucleon longitudinal momenta (the Bjorken variable for the
nucleon, $x=x_B$).

In the on-shell relativistic approach the nucleon distribution
function is connected to the well known nucleon spectral function
$S_N$ (discussed in the next section): 
\begin{equation}
\rho ^{A}(y_A) = \int d^{4}p \delta \left ( y_A/A-p^+ /P^+ \right
) S_N(p^o,{\bf p}) . \label{eq:rho}
\end{equation}
In the mean field approximation $S_N(p^o,{\bf p})=n(p)\delta
(p_o-(m+e(p))$, where $e(p)$ is the nucleon single particle energy.
This expression should be corrected by the incident flux factor (see
eq.(\ref{eq:flux}). It turns out, however, that even then one cannot
describe the EMC data without inclusion of higher order nucleon-nucleon
correlations (with additional free parameters and all uncertainties of
off shell behavior of nucleons it brings in) \cite{Ben}.

\section{Nuclear Relativistic Mean Field.}

In the nuclear relativistic mean field (RMF) method electrons collide
with nucleons which are moving in some constant average scalar and
vector potentials in the rest frame of the nucleus according to the
 equation: 
\begin{equation}
[{\hbox{$\boldmath{\alpha}$}} \cdot {\bf p} + \beta (m+U_{S}) -
(e_{N}-U_{V})]\psi = 0 . \label{eq:Dirac}
\end{equation}
Here $U_{S}= -g^{2}_{s}/m^{2}_{s} \rho _{s},~U_{V}=V_{\mu }\delta
_{\mu 0}= g^{2}_{v}/m^{2}_{v} \rho $ with $g_i,\ m_i\ (i=s,v)$ being
the scalar or vector meson coupling constants and their masses,
respectively, whereas $\rho_{s}= \sum^{}_{i}\psi ^{+}_{i}\beta \psi
_{i}$ and $\rho = \sum^{}_{i}\psi ^{+}_{i}\psi _{i}$ are the scalar
and the fourth component vector densities, respectively. The scalar and
vector mean fields  were  investigated successfully in nuclear Dirac
phenomenology with the values $U_{V}=300, U_{S}=-400$ [MeV$(\rho
/\rho _{0})]$ $\rho_{0}=0.17fm^{-3}$ \cite{Dirac-appli}. It turns out
that $U_S$ and  $U_V$ usually cancel each other in the energy or
external response functions but their relatively big values can
explain the enhancement of the spin-orbit part of nucleon-nucleus
optical potential \footnote{One of the advantages of RMF approach is the
equation of state it leads to. For example, in Walecka model
\cite{Walecka} it gives properly the saturation point for nuclear
matter and no density saturation for neutron matter.}.
It was shown \cite{Shown} that $\rho^A(y_A)$ depends on both the
scalar and vector nuclear fields:
\begin{equation}
 \rho ^{A}(y_A) = {4\over \rho }\int {d^{4}p\over
(2\pi)^{4}} S_N(p^o,{\bf p})(1+p^*_3/ E^{^{*}}(p))\delta
(y-(p_o+p^*_3)/\mu), \label{eq:flux}
\end{equation}
where  $\mu$ is equal to the nucleon chemical potential, factor
$(1+p^*_3/ E^{^{*}}(p))$ corrects (\ref{eq:rho}) for relativistic
effects and the nucleon spectral function is taken in the RMF
approach to be equal $S_N=n(p)\delta ( p^o-(E^{^{*}}(p) +U_V)) $. Eq.
(\ref{eq:flux}) can be simplified and written as:
\begin{equation}
\hspace{-3mm} \rho ^{A}(y_{A}) = {3\over 4}
(v^{2}_{A}-(y_{A}-1)^{2})/v^{3}_{A}, \label{eq:sinply}
\end{equation}
where $v_{F} =
(\mu/m)(p_{F}/E_{F}^{^{*}})$,~$v_{A}=p_{F}/E^{^{*}}_{F}$ and $y$
is restricted to region $0<(E_{F}^{^{*}}
-p_{F})<my<(E_{F}^{^{*}}+p_{F})$. It means that all nuclear
dependence is hidden in the nucleon chemical potential, which is,
however, too weak (about $8$ MeV smaller than $m$) to reproduce the
minimu seen in the EMC data \cite{DATA} at $x=0.7$ (cf. curve $(a)$
in Fig. 2). 

\section{Proposed model for parton distribution in nuclei}

Let $j$ denote four-momenta of struck parton (probed by current
with virtuality $Q^2_0$) selected (for valence quarks) from Gaussian
primordial distribution with width $0.172$ GeV, and $r$ the
respective four-momentum of hadronic remnants. Let also $W$ 
and $W'$ denote their respective invariant masses. Events are
accepted if:  
\begin{equation}
0 \leq j^2 \leq W^2 , 0 \leq r^2 \leq W'^2 \label{eq:boundary}
\end{equation}
The sea parton distribution is given by the convolution of the pionic
component of the nucleon, $f_\pi(x;Q^2_0)$, and the parton structure
of pion, $f_{pion}(x;Q^2_0)$, obtained from the same Gaussian
primodial distribution as used for valence partons\footnote{Actually,
in this way we obtain the light cone target rest frame variable
$x=x_{LC}=k^+/p^+$ (where $p =(M,0,0,0)$) with a fixed resolution 
$Q^2=Q_0^2$, whereas experimentally accessible is Bjorken variable
$x=x_{Bj}=Q^2/(2 p\cdot q)$. However, in the ratio presented in Fig. 2,
the corrections introduced when processing from one $x$ to another
cancell and therefore the experimental $x$ in Fig. 2 will be
identified with our $x$.},
\begin{equation}
f_s(x;Q^2_0)=\int \frac{dy}{y} f_{\pi}(y,Q^2_0)
f_{pion}(x/y;Q^2_0). \label{eq:sea}
\end{equation}
\vspace{.cm}
\hspace{-1.cm}
\begin{figure}[h]
\begin{minipage}[t]{0.475\linewidth}
\centering
\includegraphics[height=6.cm,width=6.cm]{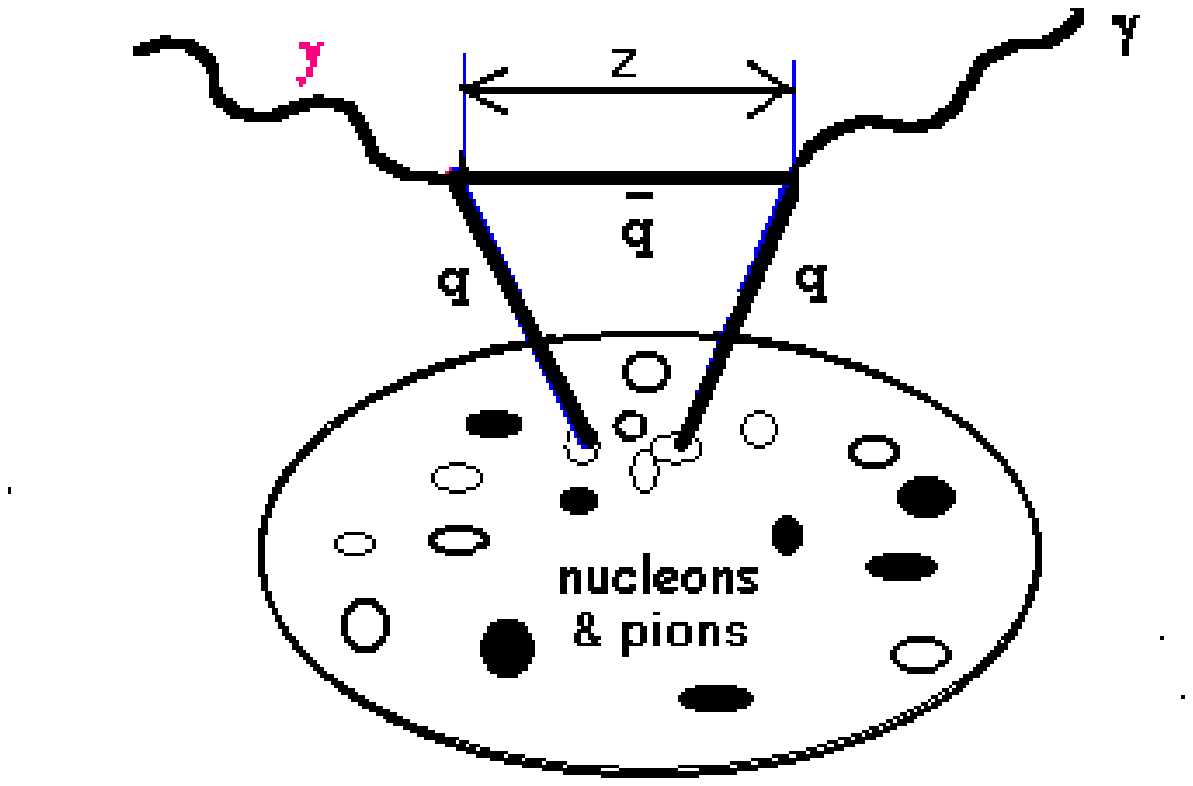}
\caption{}
\end{minipage}\hspace{-1mm}
\begin{minipage}[t]{0.475\linewidth}
\centering
\vspace{-78mm}
\includegraphics[height=9.cm,width=8.cm,angle=270]{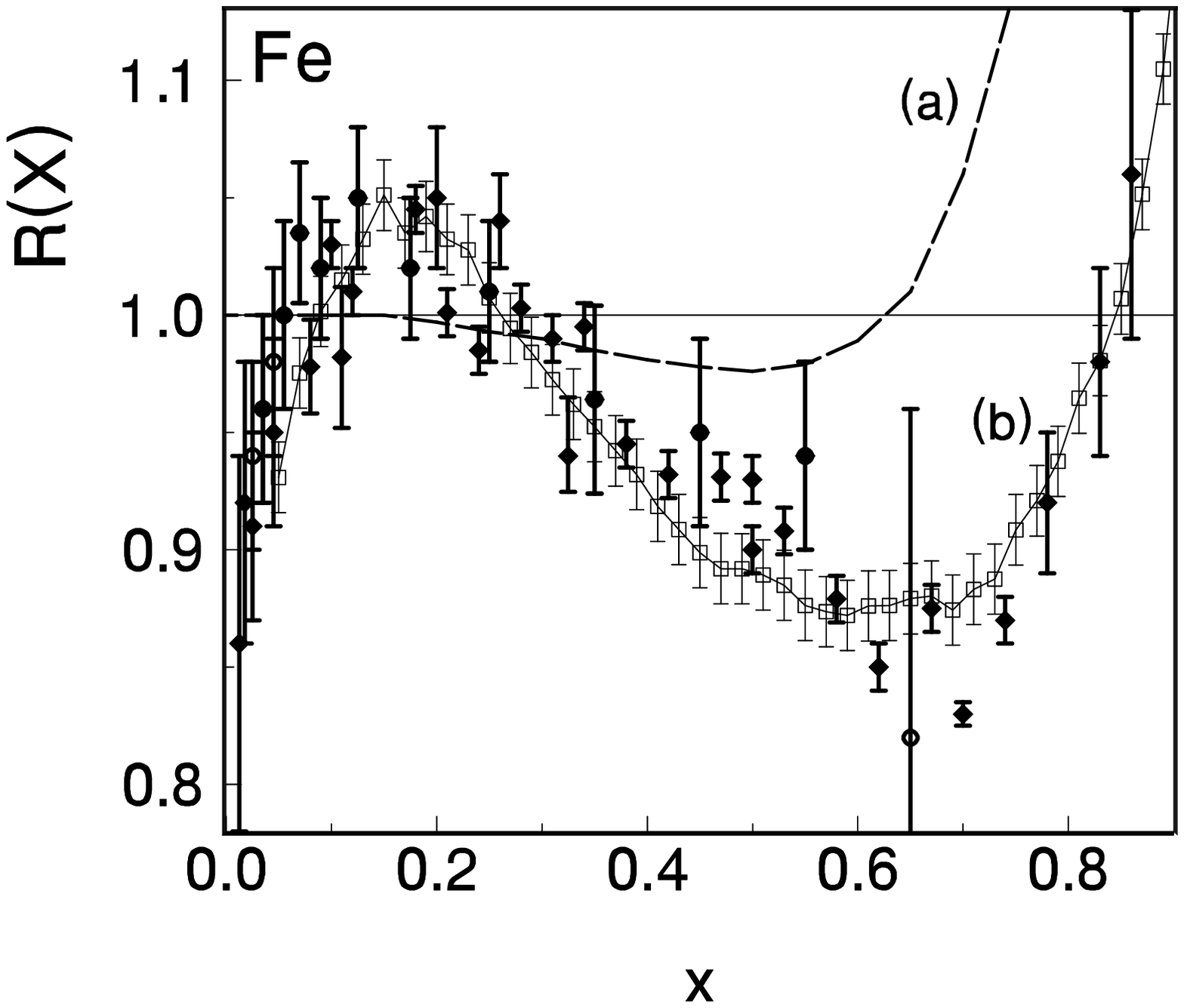}
\caption{}
\end{minipage}
\end{figure}

The characteristic behaviour of the sea partons is  derived from
the pion distribution in the nucleon, which was again parametrized
by Gaussian distribution, this time with a smaller width equal to
$0.052$ GeV. The overall partonic distribution is therefore given by
($Q_0^2 = 1$ GeV$^2$)
\begin{equation}
F_2(x;Q^2_0) = f_v(x,Q^2_0)+f_s(x;Q^2_0). \label{eq:res}
\end{equation}

Our results for R(x)=$F_2^{Fe}$/$F_2^{D}$ are presented with 
Monte Carlo error as (b) in Fig.2. 
For small $x$ the crucial factor turns out to be the change of
nuclear virtual pion cloud connected with exchanged mesons
responsible for the the nuclear forces \cite{RW}. In order to be able
to fit data in this region we have to adjust in our model the value
of the parameter which determines the relative number of the
(effective) intermediate pions (assumed to mediate nucleon-nucleon
interaction). In our model only part of pions contribute to the
sea quark structure function of the nucleon whereas the other part is
responsible for the the nucleon-nucleon interaction. Because for
small $x$ the scale $z$ shown in Fig. 1 becomes comparable or bigger
then the nucleon size, the expected range of nuclear forces also
grows accordingly. Therefore for small values of $x$ the part of
pionic contribution can "disappear" during the interaction with the
electromagnetic probe and consequently gives no contribution to the
nuclear structure function. In our case up to $12\%$ of pions are
excluded from interaction by this mechanism. 

For intermediate $x$ ($\sim 0.7$) the minimum in the EMC ratio is
obtain by adjusting the nucleon size in the medium (strictly
speaking the size of the valence parton momentum distribution).
The corresponding increase of this width from 0.18GeV to 
.172GeV  produces both the minimum for $\sim 0.7$ and the maximum
for $x$ around $0.1$. In hadronic language this change corresponds to
some spreading of the pionic cloud outside the nucleon\footnote{Cf.
the corresponding calculations performed for the realistic nuclear
distributions with momentum distribution $n(p)$ taking into account
the long tail obtained from the nucleon-nucleon residual interaction
in the mean field: $n(p)=n_{mf}(p)+n_{tail}$ \cite{Zab}.}.

\section{Summary}

We obtain very good fit to the data on deep inelastic
scatterings with $^{56}Fe$  using only two physically motivated
parameters\footnote{Similar results for the broad range of $A$ will
be presented elsewhere.}. The first describes decrease of width of 
the primordial gaussian valence quark
distributions (it points towards the possible deconfinement of quarks
in nuclear matter and to chiral symmetry restoration). The second
parameter diminish the  amount of nuclear pions  below $x=0.1$ due
to the
shadowing effect.

\end{document}